\documentclass[twocolumn,showpacs,preprintnumbers,pre]{revtex4}

\usepackage{amssymb}
\usepackage{amsmath}
\usepackage{graphicx}
\usepackage{psfrag}
\usepackage{subfigure}
\usepackage{epsfig}

\begin{document}

\title{Finite-distance singularities in the tearing of thin sheets}

\author{E. Bayart, A. Boudaoud\dag, and M. Adda-Bedia}

\affiliation{Laboratoire de Physique Statistique, Ecole Normale Sup\'erieure, UPMC  
Paris 06, Universit\'e Paris Diderot, CNRS, 24 rue Lhomond, 75005 Paris,  France.}

\date{\today}

\begin{abstract}
We investigate the interaction between two cracks propagating in a thin sheet. Two different experimental geometries allow us to tear sheets by imposing an out-of-plane shear loading. We find that two tears converge along self-similar paths and annihilate each other. These finite-distance singularities display geometry-dependent similarity exponents, which we retrieve using scaling arguments based on a balance between the stretching and the bending of the sheet close to the tips of the cracks.
\end{abstract}


\maketitle
Thin sheets are widespread in Nature and technology. Examples include insect wings, leaves or tectonic plates, and graphene, conducting layers or metallic roofs, respectively. Strikingly, these objects feature two types of energy focusing, at a sharp fold as in crumpled paper~\cite{witten} or at the tip of a crack as in a torn sheet of paper. More generally, understanding and predicting the propagation of a crack in a brittle material yield a central challenge in fracture mechanics~\cite{freund,leblond}. A body of research has been performed on the path of a crack submitted to in-plane tensile (mode I) or shear (mode II) loading, see~\cite{katzav07,corson09,bouchbinder10} for recent references. An emerging consensus is that a crack generally propagates along a path such that mode II vanishes and the material opens in a tensile mode (principle of local symmetry). Few studies considered out-of-plane shear (mode III) loading, see~\cite{barenblatt61,lazarus08}. Even fewer investigations addressed mode III crack propagation in thin sheets, focussing on ductile materials~\cite{mai84,atkins94,yan09}, on sheets adhering to a substrate~\cite{hamm08,sen10}, or on cutting with blunt indentors~\cite{atkins94,roman03,ghatak03}. This class of experiments point out the coupling between the bending of the sheet and in-plane stretching leading to the opening of cracks.

In a preliminary report~\cite{bayart10}, we studied the propagation of one or two cracks in a brittle thin sheet, using the so-called trousers test configuration~\cite{mai84}. The results that we obtained indicate that, although a large scale mode III is imposed, the material locally breaks in the tensile mode I. As a consequence, the interfaces opened by the crack are not perpendicular to the sheet surface and are inclined by 45 degrees instead. Here we also consider a peel-like configuration and we focus on two tears propagating quasi-statically. We find that they converge along self-similar paths, with characteristic exponents corresponding to each type of loading. The topological change that occurs when they annihilate each other is reminiscent of finite-time singularities in the breakup of liquid droplets and jets~\cite{eggers97,eggers08}, which suggests the terminology of finite-distance singularity. Our power-law crack paths differ from the exponential shapes predicted in~\cite{cohen10};  we account for the exponents that we observe using scaling arguments inspired by energy localization along ridges in crumpled sheets~\cite{lobkovsky97}. Therefore our experiments appear as a non-trivial combination of the two types of energy focusing known in thin sheets.

In experiments, we used thin films of bidirectional polypropylene of thickness $h=$ 30, 50 and 90$\mu$m. The Young's modulus and fracture toughness were measured as $2.2\pm0.4$ GPa and $2.6\pm0.3$MPa.m$^{1/2}$, respectively. The films were found to have isotropic mechanical properties within these uncertainties. All experiments were performed quasi-statically, with crack velocities in the range 0.05--1.5mm.s$^{-1}$. At ambient temperature and in this velocity range, the fracture process is brittle for polypropylene. The crack paths were digitized using a scanner, with a resolution of about 10$\mu$m.

\begin{figure}
\centering
\begin{minipage}[c]{0.55\columnwidth}
a\\
\includegraphics[width =\textwidth]{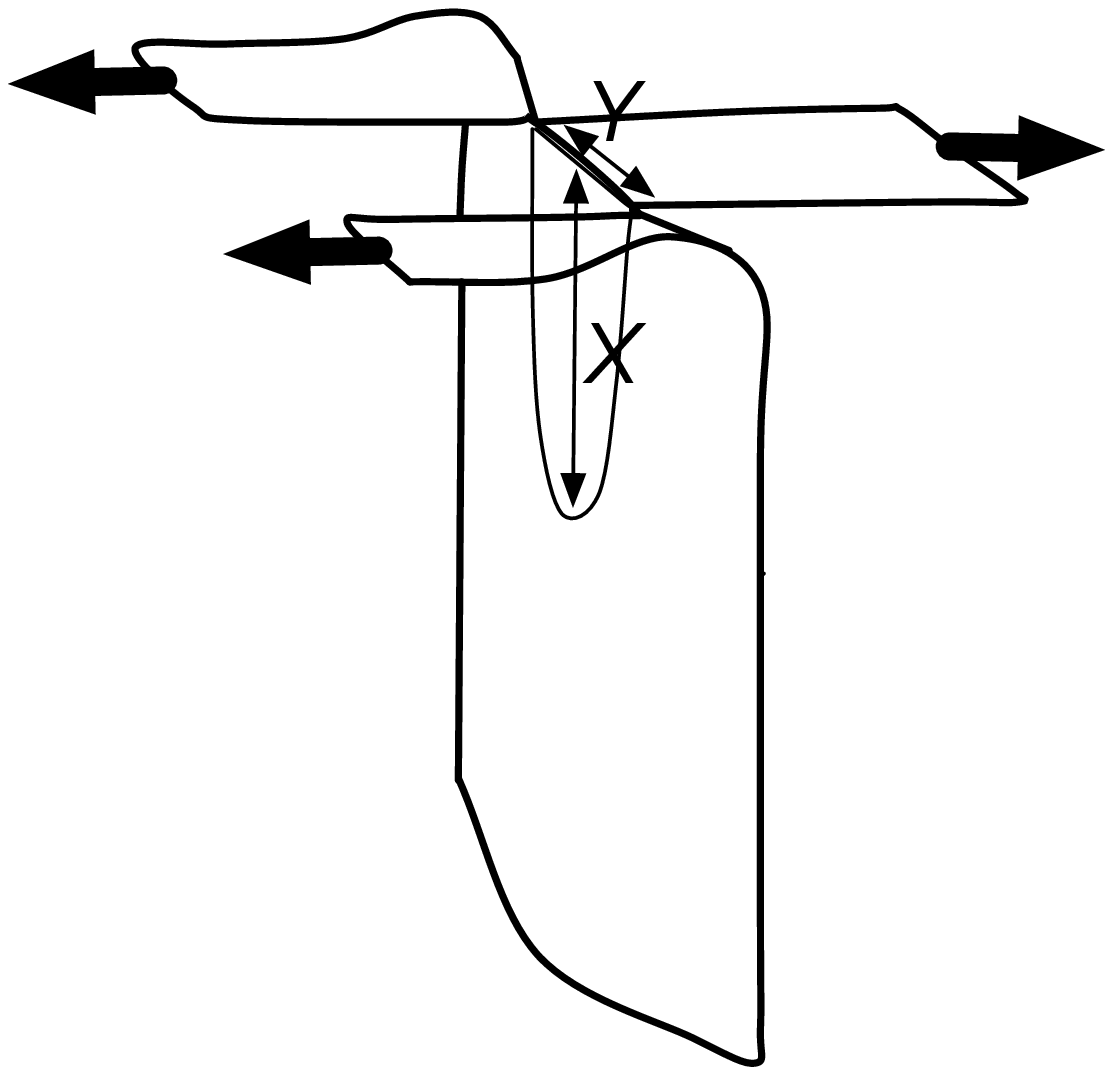}
b\\
\includegraphics[width =\textwidth]{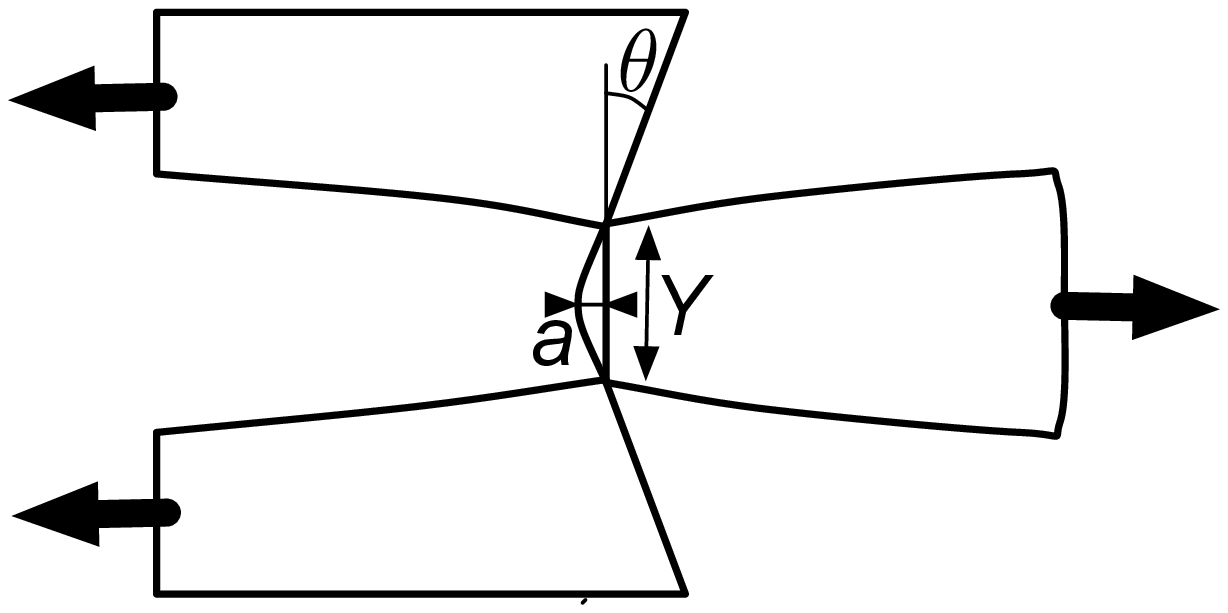}
\end{minipage}
\begin{minipage}[c]{0.43\columnwidth}
c\\
\includegraphics[width =\textwidth]{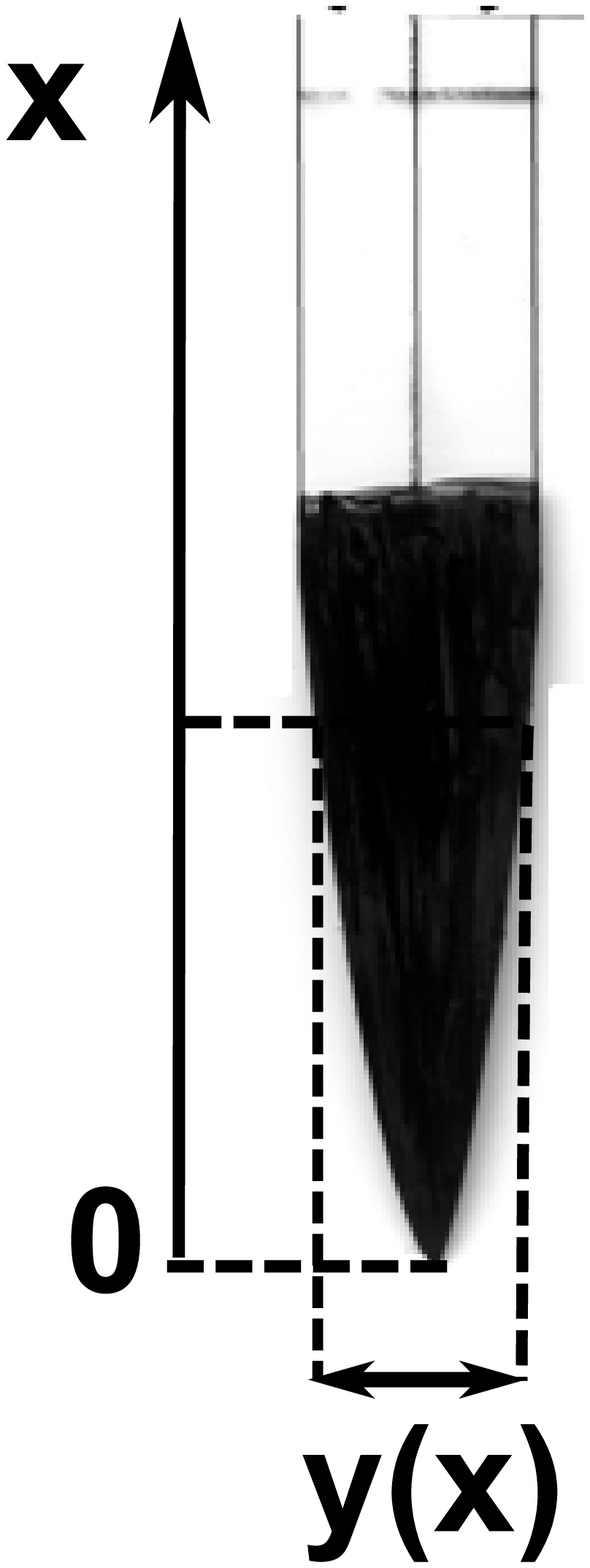}
\end{minipage}
\caption{The trousers configuration: side (a) and top (b) schematics of the experiment. The sheet is pulled from the three strips as shown by the arrows. The strips are deflected by freely rotating cylinders and then rolled over two cylinders entrained by a continuous motor. (c) After the tears have converged, the central strip has a tongue-like shape, described by a curve $y(x)$.}
\label{fig1}
\end{figure}

The first setup is inspired by the trousers test used to characterize ductile sheets~\cite{mai84} and mimics the common way one tears a sheet of paper in three pieces. It was described in~\cite{bayart10}; we recall here the important features of the setup. In a very long sheet of width $W$, two cracks are initiated by two cuts positioned symmetrically with respect to the central axis of the sample, so that three strips are created at one end of the sheet (Fig.~\ref{fig1}).  The sheet is pulled from the three strips using cylinders for the entrainment. The main control parameters are the width of the sample $W$, the initial distance between the two cracks $w$, and the distance $d$ between the freely rotating cylinders that deflect the strips. We found that the results presented here are independent of $d$. When the tears are propagated the distance between their tips, $Y$, decreases from $w$ to 0, as they converge and eventually annihilate each other. The topology has changed: the sheet is split in two parts, and the central strip has detached into a tongue-like shape, as shown in~Fig.~\ref{fig1}c. When the ratio between the initial width of the central strip and the width of the sample is smaller enough $(w\lesssim0.1\,W)$, experiments are reproducible and the results are  found to be independent of $W$. If the central strip is larger $(w\gtrsim0.1\,W)$, experiments are less reproducible and the shape of the tear is not always symmetric. This might be ascribed to the sensitivity of propagation to imperfections in the parallelism of the cylinders' axes.

\begin{figure}
\centering
a\\
\includegraphics[width =0.43\textwidth]{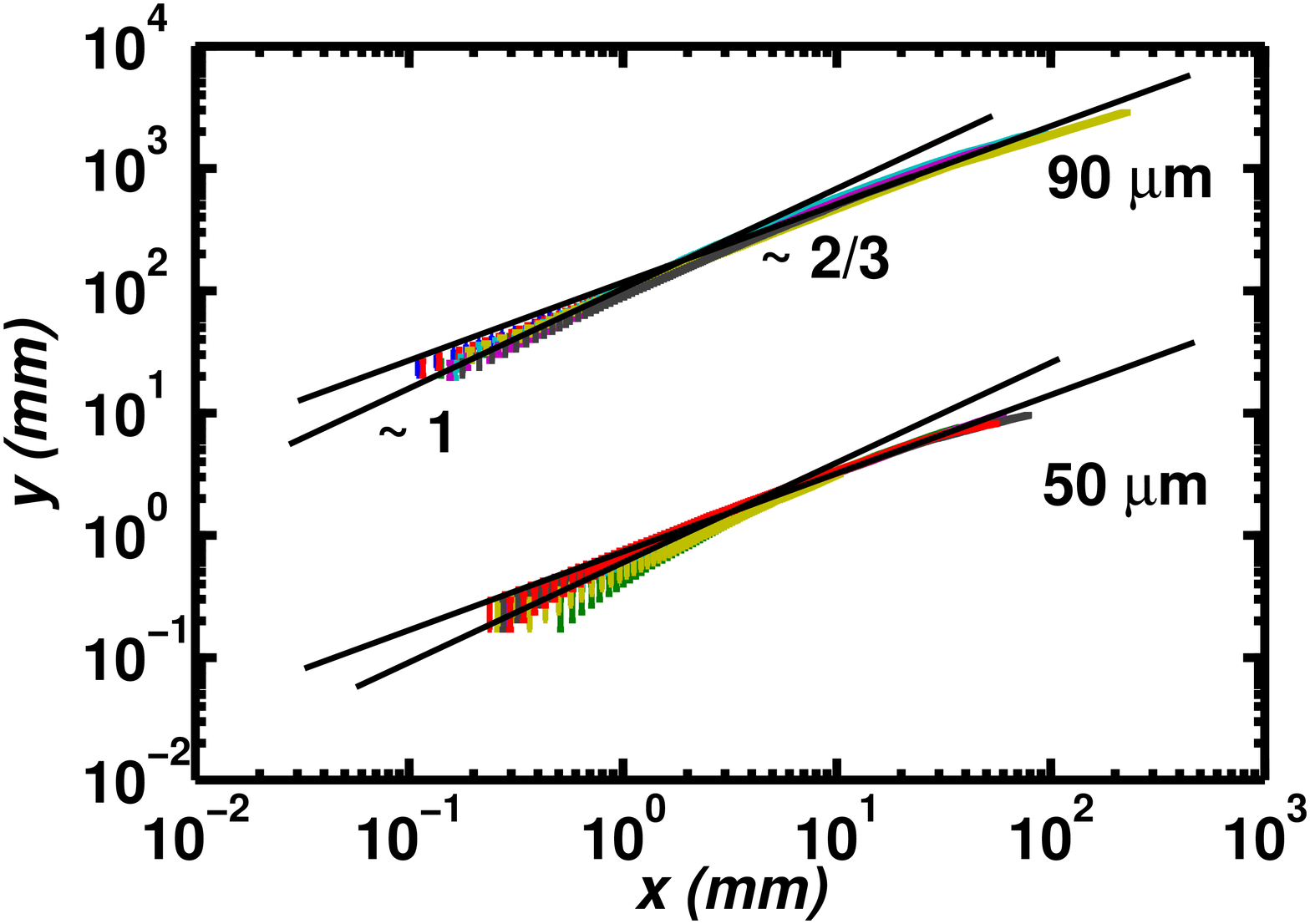}\\
b\\
\includegraphics[width =0.43\textwidth]{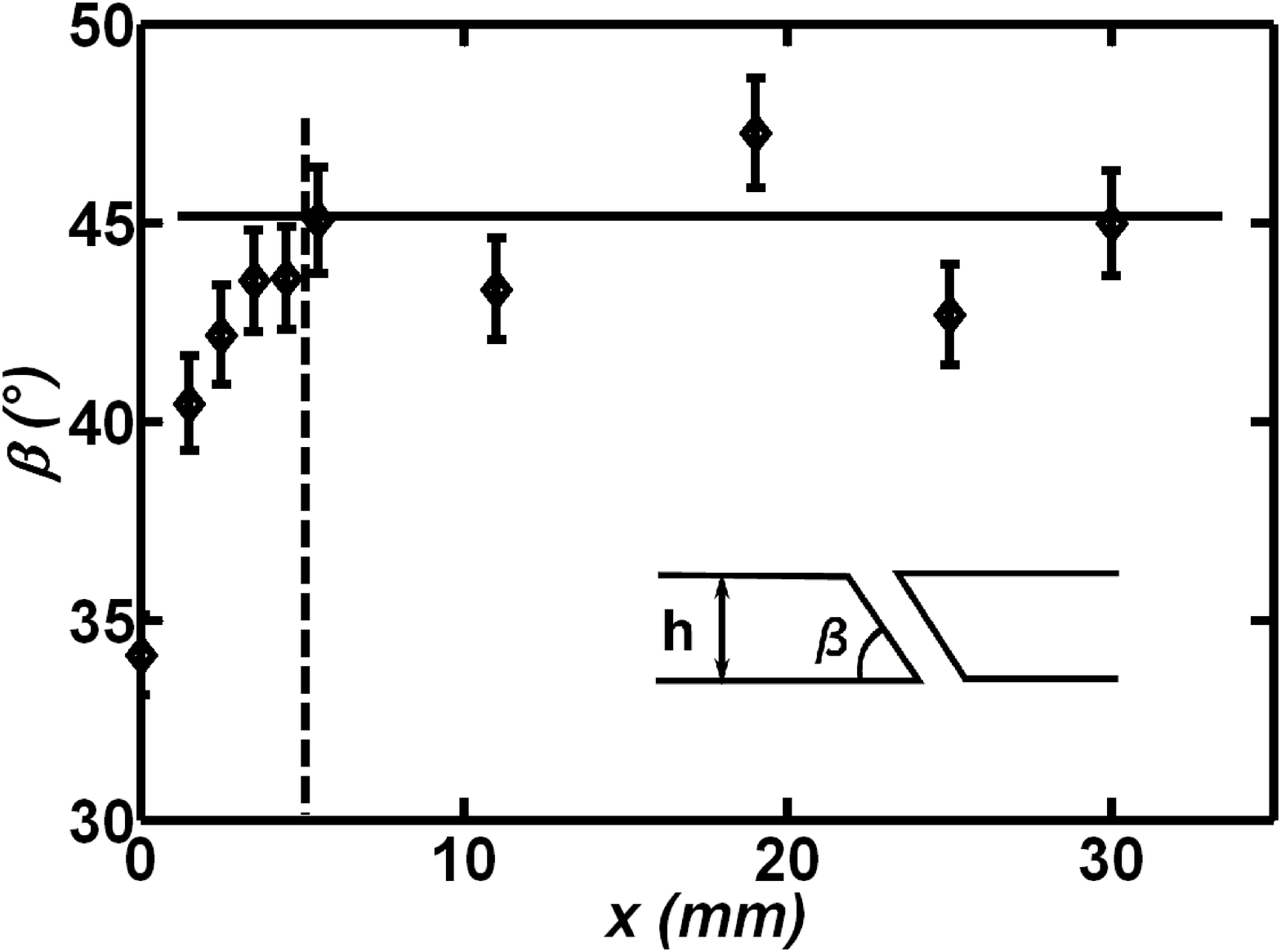}\\
c\\
\includegraphics[width =0.43\textwidth]{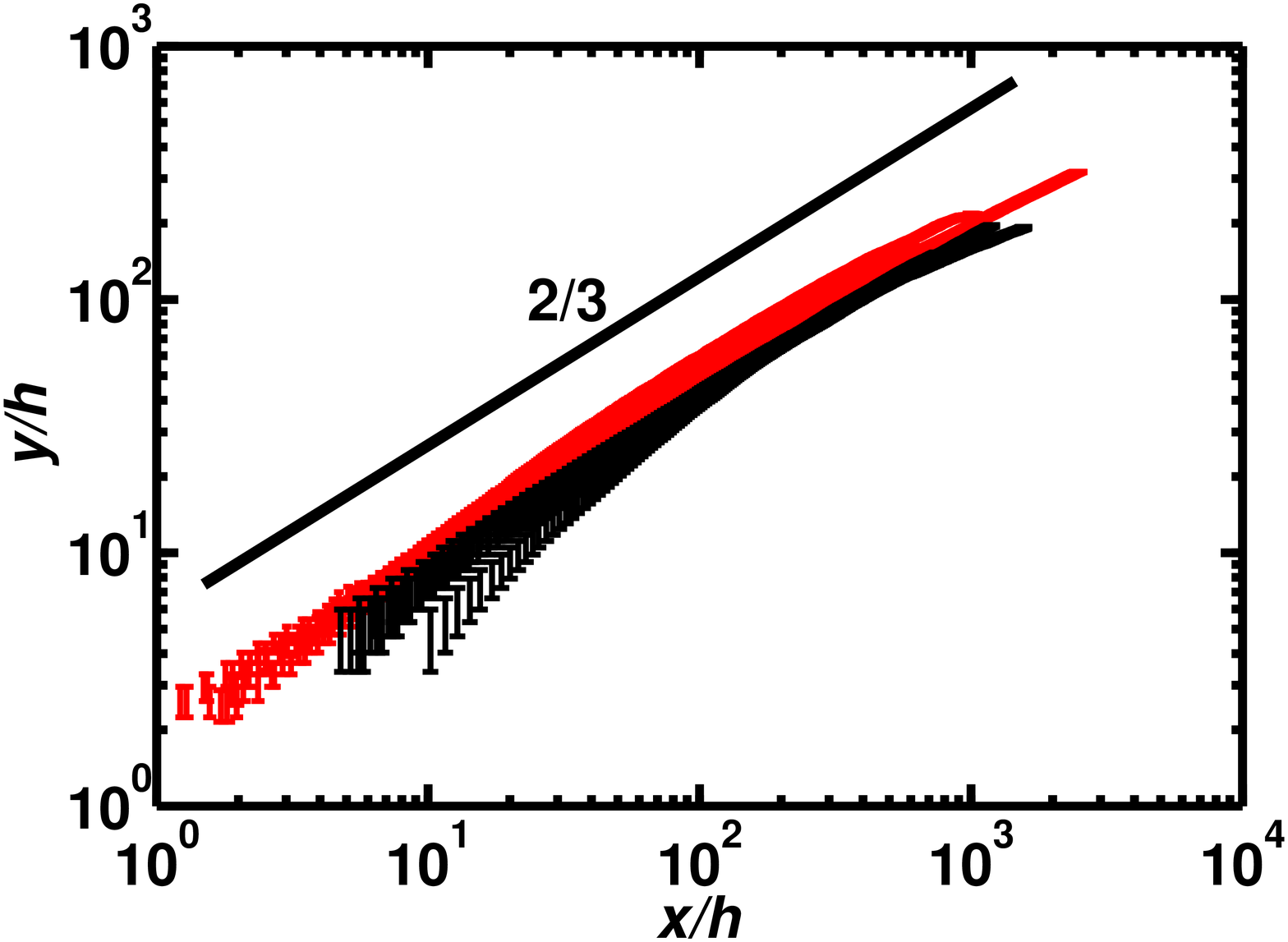}
\caption{Crack paths in the trousers configuration. (a) Shape of the central strip: width $y$ as a function of longitudinal coordinate $x$, see Fig.~\ref{fig1}c for definitions. The widths for a thickness of 90$\mu$m were multiplied by 100 for clarity. The colors of the symbols correspond to different realizations. Power laws of exponents 2/3 and 1 are shown for comparison. (b) Angle $\beta$ of inclination of the fracture surface with respect to the sheet surface (defined in inset) as a function of the longitudinal distance $x$, suggesting a transition around $x=5$mm from 45$^\circ$ to lower values. The initial distance between the two tears was $w=20$mm and the thickness had the value $h=50\mu$m. (c) The curves in (a) were made nondimensional using the sheet thickness $h$ as a unit. Black and red symbols correspond to $h=50\mu$m and $90\mu$m, respectively}
\label{fig2}
\end{figure}

We analyzed the crack paths by digitizing the tongue-like central strip. The width $y$ of the strip is defined in Fig.~\ref{fig1}c and shown in Fig.~\ref{fig2}a as a function of the distance $x$ to the point where the two tears converged. An important observation is that all realizations are superimposed (for various values of $w$, $W$, and $d$), indicating that the geometry of the setup is unimportant for path selection. We found that, in the range $5\leq x\leq300$mm, the curves $y(x)$ are well-described by a power-law of exponent $0.64\pm0.06$ and a prefactor ($0.77\pm0.2$ and $1.2\pm0.2$) that is larger for the largest value of thickness. Below $x=5$mm, $y(x)$ seems close to being linear. We then sought the origin of this transition and made the following observations. After tearing, far from the tip, the strip recovers its flatness indicating a brittle fracture process, while it is permanently curved in a small region close to the tip, which is a signature of plastic deformations. Post-mortem analysis of the fracture surface shows that it is generally inclined by an angle $\beta=45^{\circ}$ with respect to the surface of the film, in agreement with our previous observations on a single crack~\cite{bayart10}. However, a transition occurs for $x\sim5$mm: this angle decreases down to the value $\beta=34^{\circ}$ at the merging point. This transition seems to be related to the appearance of plastic deformations. If $\beta$ decreased to zero, the fracture surfaces would be parallel to the sheet surfaces, and the configuration would resemble the peeling of an adhesive strip, in which two notches lead to triangular strips~\cite{hamm08}. This remark might account for a linear behavior of $y(x)$ for small $x$. Finally, we non-dimensionalized the profile $y(x)$ by using the thickness $h$ as a unit (Fig.~\ref{fig2}c), which yields a fair collapse of the data. However, we were unable to obtain reproducible results with the smallest thickness, $h=30\mu$m, for which the crack paths are very sensitive to the forcing. As a consequence, the range in thickness is too narrow to rule out other small-scale characteristic lengths. 

\begin{figure}
\centering
a\\
\includegraphics[width =0.45\textwidth]{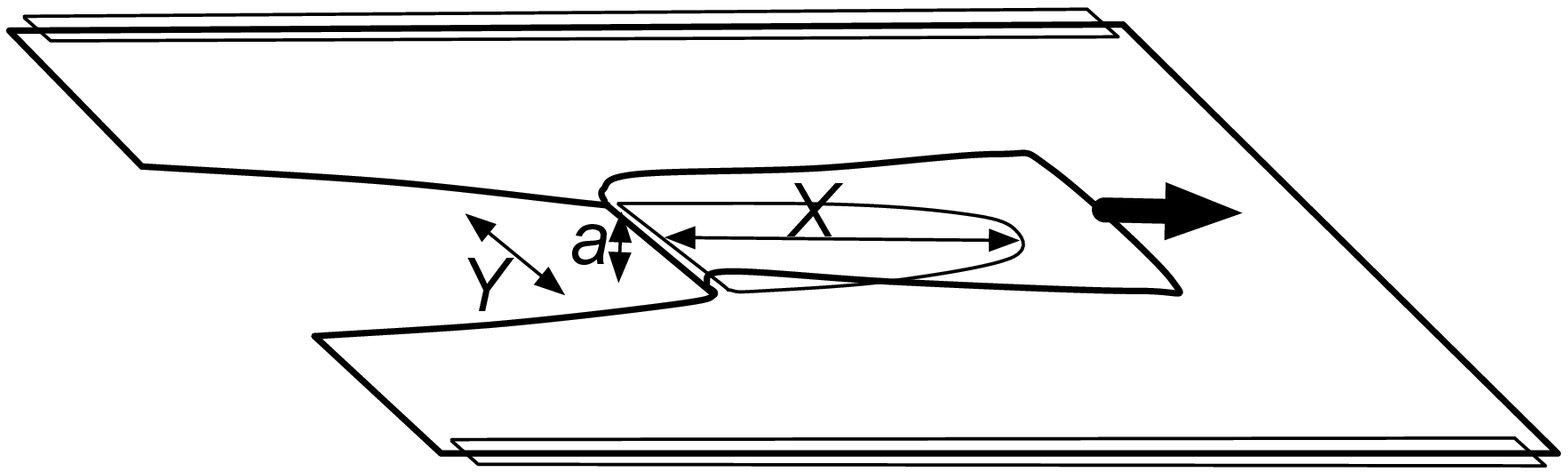}\\
b\\
\hspace{.2\columnwidth}\includegraphics[width =0.6\columnwidth]{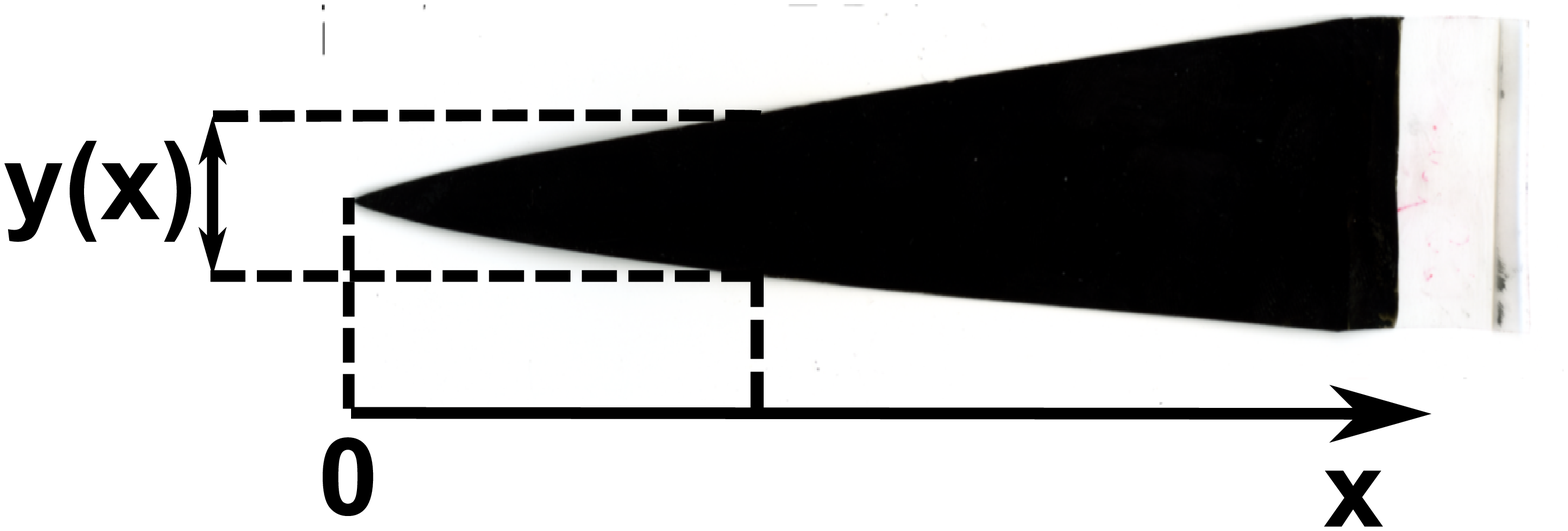}
\caption{The peel-like configuration. (a) Schematic of the experiment. Two notches are initiated in a rectangular long sheet clamped along its lateral boundaries. The central strip is pulled horizontally at constant velocity until it detaches when the two tears converge. (b) Tongue-like shape of the detached strip, defined by $y(x)$. }
\label{fig3}
\end{figure}

\begin{figure}
\centering
a\\
\includegraphics[width =0.43\textwidth]{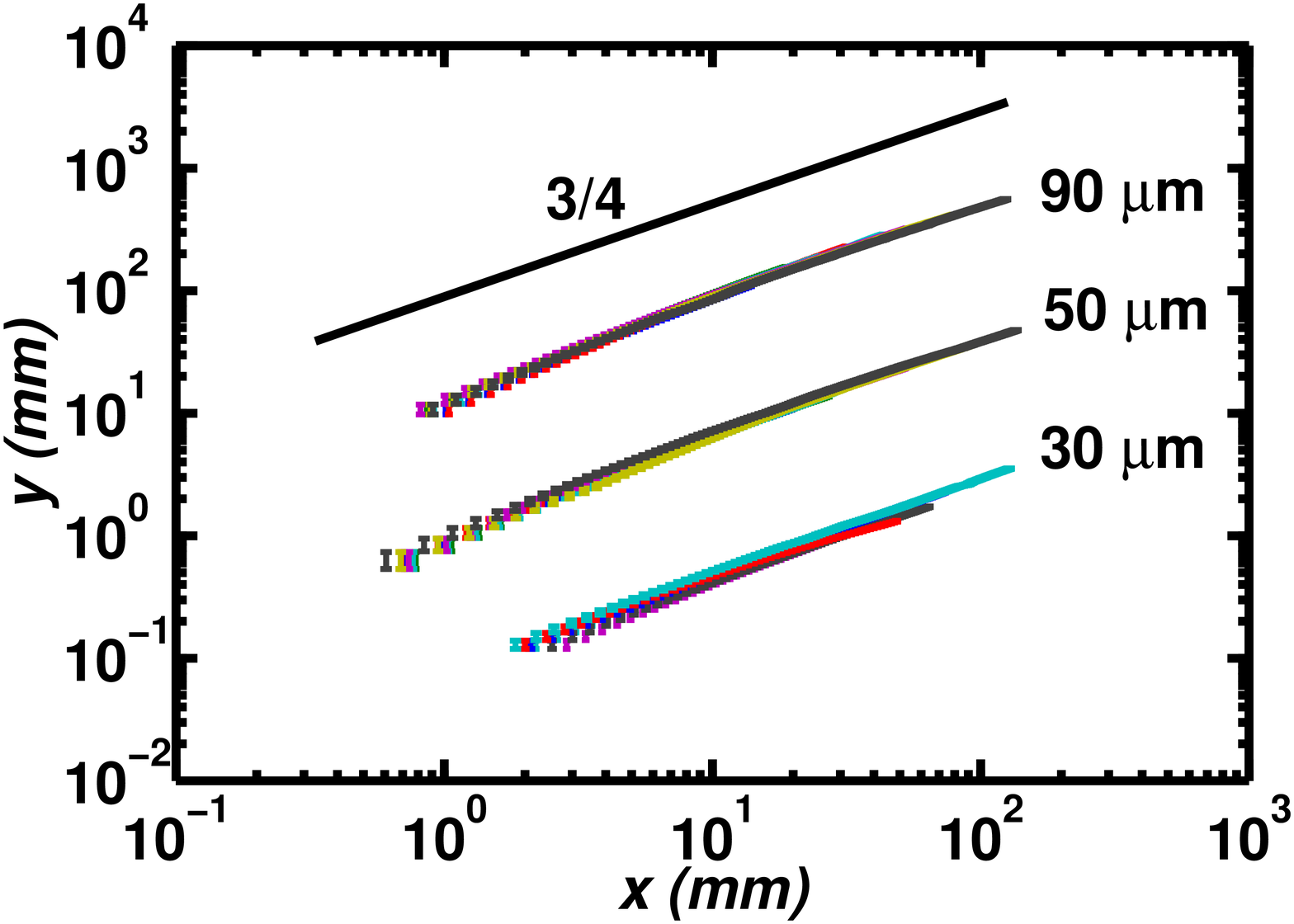}\\
b\\
\includegraphics[width =0.43\textwidth]{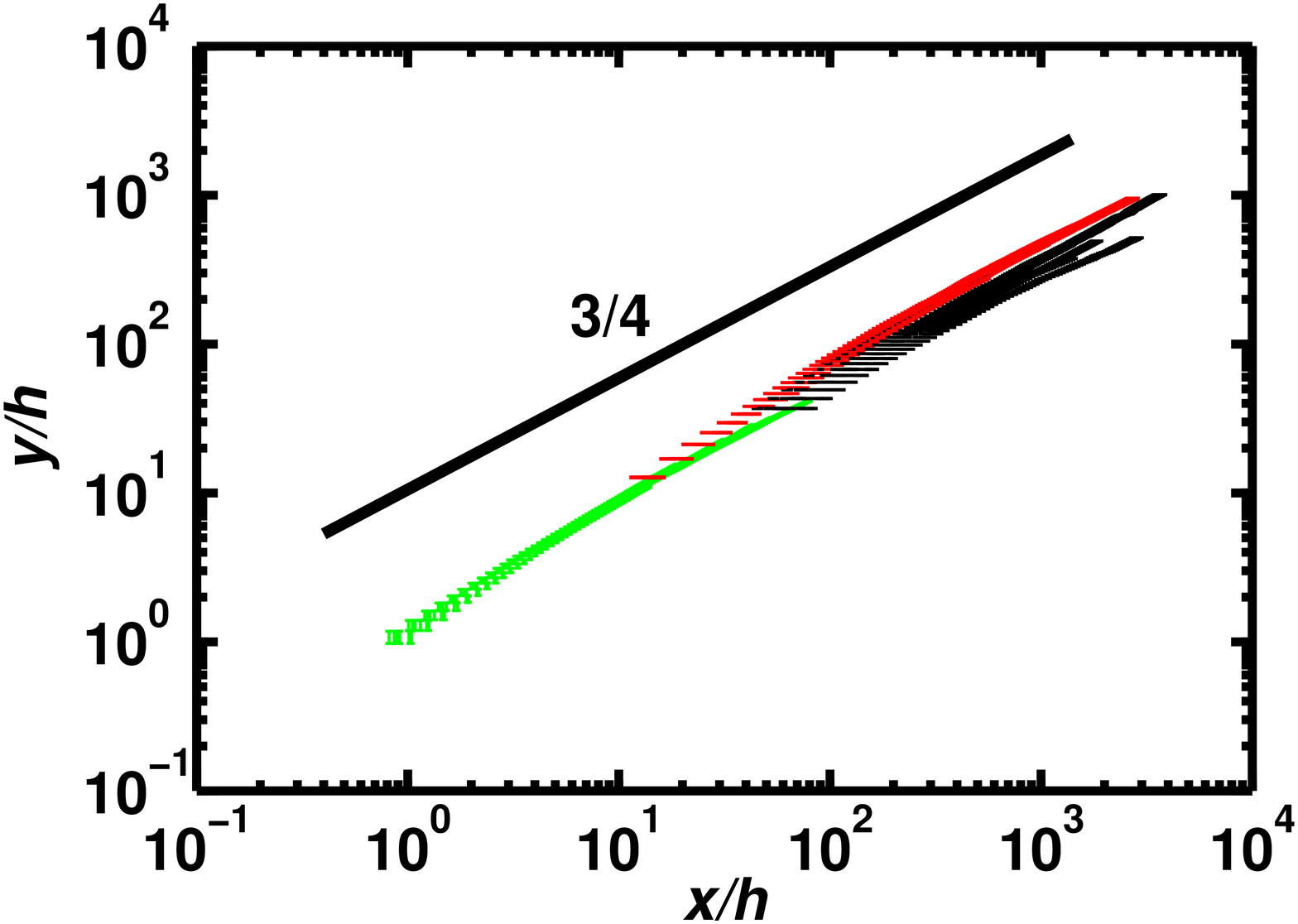}
\caption{Crack paths in the peel-like configuration. (a) Shape of the central strip: width $y$ as a function of longitudinal coordinate $x$, see Fig.~\ref{fig3}b for definitions. The widths for the values of the thickness 30$\mu$m and 90$\mu$m were divided and multiplied by 10, respectively, for clarity. The colors of the symbols correspond to different realizations. A power laws of exponents 3/4 is shown for comparison. (b) The curves in (a) were made nondimensional using the sheet thickness $h$ as a unit. Green, black and red symbols correspond to $h=30\mu$m, $50\mu$m, and $90\mu$m, respectively}
\label{fig4}
\end{figure}

In a second step, we considered a less symmetric, peel-like experimental configuration, which was inspired from~\cite{hamm08}. A long sheet of width $W$ is clamped along its lateral boundaries to a thick wooden plate, using narrow adhesive tapes (Fig.~\ref{fig3}a). Two parrallel notches are initially made at a distance $w$ one from an another at one end of the sheet. The central strip is pulled horizontally, so that the two tears propagate quasi-statically. In contrast with the first configuration, the distance between the pulling point and the crack tips increases, but this macroscopic length appears to be unimportant in the following results. As the tears advance, the distance between the two tips decreases from $w$ to 0, when they annihilate each other and the central strip detaches. The resulting shape $y(x)$ (Fig.~\ref{fig3}b) is qualitatively similar to those of the first setup. The shapes of various detached strips is shown in Fig.~\ref{fig4}a. Again, all realizations are superimposed for a given value of thickness, indicating that the geometry of the setup is unimportant for path selection. Over two orders of magnitude, the curves $y(x)$ are well described by a power-law $0.77\pm0.05$ and a prefactor ($0.82\pm0.06$, $1.2\pm0.2$, and $1.53\pm0.2$) that increases with thickness. As in the first setup, the tip of the central strip undergoes plastic flow, but over a smaller length ($\sim1$mm). However, we have not found any signature on the crack paths. Finally, we non-dimensionalized the profile $y(x)$ using the thickness $h$ as a unit (Fig.~\ref{fig4}b), which provides a satisfying collapse of the data.


Overall, the propagation of the two tears leads to a topological change such that the cracks annihilate each other and the central strip detaches. The crack paths seem to follow well-defined power-laws. In order to explain this behavior, we consider the ridge joining the two tips, which is one form of energy localization in thin sheets~\cite{witten} as the sheet is highly bent. Guided by the principle of maximal release of energy for the selection of a crack path, we postulate that the two cracks follow the stress field induced by the presence of this ridge and we investigate how this stress field decays away from the ridge. This ridge of length $Y$ imposes a normal displacement $a$ that is felt over a distance $X$ in the region ahead of the cracks (see Figs.~\ref{fig1}a,b and~\ref{fig3}a). Following~\cite{lobkovsky97}, we estimate the length $X$ by balancing the stretching energy $\mathcal{E}_\mathrm{s}$ and the bending energy $\mathcal{E}_\mathrm{b}$ in a region of area $S\sim XY$. Assuming $Y\ll X$, the strain $s$  is distributed over the larger dimension of this region so that it can be estimated as $s\sim (a/X)^2$, yielding a stretching energy $\mathcal{E}_\mathrm{s}\sim E h s^2 S$. The main curvature $c$ is along the shortest direction, so that $c\sim a/Y^2$, yielding $\mathcal{E}_\mathrm{b}\sim Eh^3 c^2S$. As a consequence, $\mathcal{E}_\mathrm{s}\sim\mathcal{E}_\mathrm{b}$ corresponds to 
\begin{equation}
Y\sim X(h/a)^{1/2}, 
\label{scale-a}
\end{equation}
which is consistent with the assumption that $Y\ll X$ as long as $a\gg h$. 

In the case of the trousers configuration, the normal displacement $a$ corresponds to a slope $\theta$ felt over a distance $X/2$, so that $a\sim X\theta$. In experiments, we observed this slope $\theta\sim0.1$ to be roughly constant, but we do not have a theoretical argument for its selection. Replacing in equation~(\ref{scale-a}), we obtain
\begin{equation}
Y \sim (h/\theta)^{1/3} X^{2/3}.
\end{equation}
Due to the weak dependance on $\theta$, we expect prefactor of order 1 for $Y/h\sim(X/h)^{1/3}$. Following the postulate that the cracks follow the shape of the region where the ridge is felt, this scaling should predict the paths. Indeed, it yields a good description of experimental results in Fig.~\ref{fig2}c, except for the linear behavior for small $x$.

In the peel-like configuration, the normal displacement is given by the width of the ridge boundary layer~\cite{lobkovsky97},  $a\sim h^{1/3}X^{2/3}$. A substitution in equation~(\ref{scale-a}) leads to
\begin{equation}
Y \sim h^{1/4} X^{3/4}.
\end{equation}
This is also in good agreement with the experimental crack paths shown in Fig.~\ref{fig4}b. 
 
Thus, we are able to account for the power-law behavior in experiments by estimating the decay length of the stress field penetration in the sheet. This behavior is independent of macroscopic parameters but the exponents depend on the symmetries of the experimental configuration, as in the finite-time singularities occurring in the breakup of liquid droplets and jets~\cite{eggers97,eggers08}. This similarity, such that the coordinate along the axis of the sheet replaces time, prompted us to use finite-distance singularities to describe our findings. Our results differ from the exponential shapes predicted in~\cite{cohen10}, possibly because our sheets are an order of magnitude thinner than in their simulations. Indeed, in the case of the propagation of a single crack, the numerical results of Cohen et al.~\cite{cohen10} get closer to our experiments when they decrease the thickness. Future theoretical work should address this discrepancy, as well as other features that we have not accounted for, such as the selection of the slope $\theta$ or the transition to a three-dimensional fracture. Although the coupling between bending and stretching seem to imprint the stress field and guide the tears, the principles underlying propagation of cracks in out-of-plane shear loading are still to be established.
 
We are grateful to Laurent Quartier for his help in building the experimental apparatus, and to Eugenio Hamm and Benoit Roman, Yossi Cohen and Itamar Procaccia for inspiration and discussions.

\dag Current address: RDP, ENS Lyon, 46 all\'ee d'Italie, 69007 Lyon, France

\end{document}